\documentclass[letterpaper]{article} 
\usepackage[submission]{aaai23}  
\usepackage{times}  
\usepackage{helvet}  
\usepackage{courier}  
\usepackage[hyphens]{url}  
\usepackage{cite,graphicx} 
\usepackage{natbib}  
\usepackage{caption} 
\usepackage{algorithm}
\usepackage{algorithmic}
\usepackage{comment}
\usepackage{newfloat}
\usepackage{listings}
\DeclareCaptionStyle{ruled}{labelfont=normalfont,labelsep=colon,strut=off} 
\floatstyle{ruled}
\newfloat{listing}{tb}{lst}{}
\floatname{listing}{Listing}
\usepackage{multicol}
\usepackage{tabularx}
\usepackage{booktabs}
\usepackage{textcomp}
\usepackage{xcolor}

\title{TimbreCLIP: Connecting Timbre to Text and Images}
\author {
    Nicolas Jonason,\textsuperscript{\rm 1}
    Bob L. T. Sturm, \textsuperscript{\rm 1}
}
\affiliations {
    \textsuperscript{\rm 1} KTH Royal Institute of Technology\\
    njona@kth.se, bobs@kth.se
}

\usepackage{bibentry}

\begin{document}

\maketitle

\begin{abstract}
We present work in progress on TimbreCLIP, an audio-text cross modal embedding trained on single instrument notes. We evaluate the models with a cross-modal retrieval task on synth patches. Finally, we demonstrate the application of TimbreCLIP on two tasks: text-driven audio equalization and timbre to image generation. 

\end{abstract}
\section{Introduction}

Multi-modal models such as CLIP \cite{radford_learning_2021} have provided a foundation for creative artificial intelligence (AI) applications, such as text-to-image generation models \cite{rombach_high-resolution_2022,ramesh_hierarchical_2022,crowson_vqgan-clip_2022}.
Several cross modal models connecting audio to text and video have been proposed for general audio \cite{wu_wav2clip_2022,guzhov_audioclip_2022,elizalde_clap_2022,lee_robust_2022,wu_large-scale_2022}, speech audio \cite{shih_speechclip_2022}, and music \cite{huang_mulan_2022,manco_contrastive_2022}. 
We present ongoing work on TimbreCLIP, an  audio encoder which specifically targets single instrument notes, interfacing with textual and visual modalities in CLIP space. The name of the model comes from the word timbre, which we consider roughly as the perceptual quality of a sound that is not accounted for by pitch, loudness, and duration.

We make the following contributions: 1) We present a audio-text multi-modal model trained specifically on instrument timbre. We evaluate TimbreCLIP on cross modal retrieval on a set of synthesizer patches; 2) We explore two applications built using TimbreCLIP; a) Text-driven audio equalization, i.e., automatically equalizing audio based on a text prompt; b) Timbre-to-image generation, i.e., generating images from the timbre of an input sound.

We plan to release our code and model weights in the future. Supplementary material with audio examples is available here: \url{https://aquatic-singer-b25.notion.site/TimbreCLIP-Connecting-Timbre-Text-and-Images-a39a8fab48ef47488e9e4fed23061429}

\begin{figure}
    \centering
    \includegraphics[width=\columnwidth]{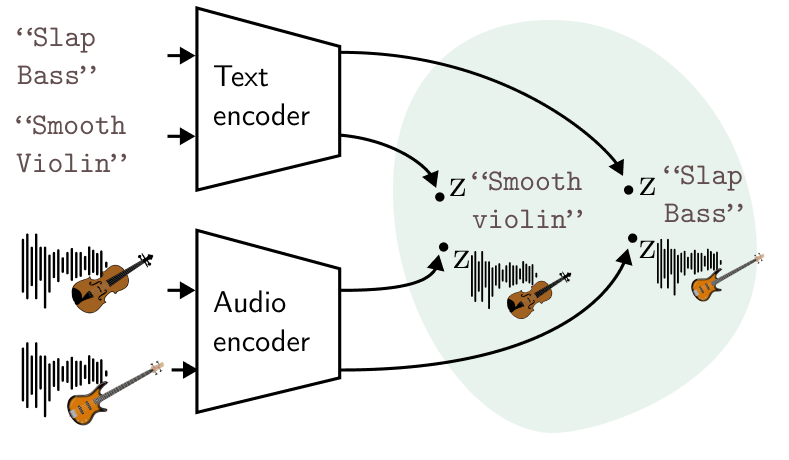}
    \caption{High level overview of how TimbreCLIP works. One encoder takes text and
    one encoder takes audio of single instrument notes. Both modalities are projected into a shared latent space. The encoders are trained such that text and audio that belong together project to points that are close in the latent space}
    \label{fig:my_label}
\end{figure}

\section{Building TimbreCLIP}

\subsection{Architecture}

Our TimbreCLIP audio encoder is trained by finetuning a Wav2CLIP model, i.e a ResNet-18 architecture pre-trained on audio-image pairs extracted from video.\cite{wu_wav2clip_2022} The text embeddings we use to train our audio encoder are produced by a frozen CLIP text encoder.\cite{radford_learning_2021} 
We use the same contrastive loss as Wav2CLIP \cite{wu_wav2clip_2022}.

\subsection{Dataset}
We perform training and model selection on a combination of two datasets. NSynth \cite{engel_neural_2017}. NSynth contains sounds from 1006 instruments with each instrument being sampled at 65.4 pitches and 4.75 unique velocities on average. Each note is held from three seconds and decays for one second. Each sound is annotated with multiple text attributes: instrument family, instrument source and notes qualities. 

The second dataset, which we will refer to as ALV, consists of 9939 patches from the commercial software synthesizer {\em Arturia} "Analog Lab V".\footnote{\url{https://www.arturia.com/products/analog-classics/v-collection/details}} We choose this software synthesizer for the following reasons: 1) it includes a wide range of patch sounds, including sounds from a variety of classic synthesizers; 2) each patch is paired with rich text annotation (title, description, multi-label tags, type and sub type). We record each patch playing the notes C and F\# for 8 octaves ranging from C0 to F\#7 totalling 16 notes per patch. All notes have MIDI velocity 64. Each note is held for three seconds and decays for one second. 

For NSynth, we use the same training/validation splits as \cite{engel_neural_2017}. For ALV, we use 90\% of the patches for training and 10\% for validation. After downsampling all audio to 16 kHz, we randomly mix the left and right channels down to mono. For the training sets, we also perform data augmentation through pitch shifting in order to provide support for pitches not present in the original datasets. Notes from the NSynth dataset are augmented with an additional pitch shifted version with offsets sampled uniformly from -0.5 to +0.5 semitones. Notes from the ALV dataset are augmented with two pitch shifted versions with offsets sampled uniformly from -3 to 3 semitones. Finally, we apply peak normalisation to all notes. We also augment the text annotation by combining various text attributes to form synthetic attributes (Details in web supplement).
\subsection{Training}
We train two models, one trained on a combination of ALV and NSynth and one trained on ALV only. We train on batches of 256 sounds each. The text used in each batch is the union of the text attributes for all the sounds in the batch. We use the Adam optimizer with a learning rate of $2e-5$ and early stopping with a patience of 10 epochs.



\section{Evaluating TimbreCLIP with Cross-Modal Retrieval}

\begin{table*}\centering
\begin{tabular}{l|rrrrr|rrrrr}
\toprule&\multicolumn{5}{c}{TEXT-TO-PATCH} & \multicolumn{5}{c}{AUDIO-TO-TEXT}\\
         model &        R@1 &        R@5 &       R@10 &       R@50 &   RANK &        R@1 &        R@5 &       R@10 &       R@50 &   RANK \\
\midrule
&\multicolumn{10}{c}{TITLE} \\
\midrule
    LAION CLAP &          0.385 &          0.743 &          0.964 &          3.313 &          899.5    &     0.114 &          0.498 &          0.972 &          4.145 &          808.6 \\
  TimbreCLIP * &          0.294 &          0.764 &          1.553 &          6.867 & \textbf{723.6} & \textbf{0.588} & \textbf{1.506} & \textbf{2.753} & \textbf{9.801} & \textbf{654.0} \\
TimbreCLIP ALV & \textbf{0.541} & \textbf{1.078} & \textbf{2.042} & \textbf{7.761} &          724.9 &          0.408 &          1.230 &          2.147 &          8.871 &          681.7 \\
      Wav2CLIP &          0.193 &          0.303 &          0.716 &          2.574 &          987.1 &          0.270 &          0.714 &          1.086 &          4.169 &          855.4 \\
       perfect &         98.672 &         99.888 &         99.967 &        100.000 &            1.0 &        100.000 &        100.000 &        100.000 &        100.000 &            1.0 \\
        random &          0.091 &          0.281 &          0.530 &          2.435 &         1009.8 &          0.115 &          0.341 &          0.617 &          2.809 &          908.8 \\
\midrule
&\multicolumn{10}{c}{TITLE + CATEGORY} \\
\midrule

    LAION CLAP & \textbf{0.292} & \textbf{0.740} & \textbf{1.413} &          4.257 &          926.9 &          0.144 &          0.450 &          0.834 &          3.485 &          811.6 \\
  TimbreCLIP * &          0.140 &          0.391 &          1.167 & \textbf{4.830} &          883.8 & \textbf{0.708} & \textbf{1.458} & \textbf{2.177} & \textbf{6.496} & \textbf{766.5} \\
TimbreCLIP ALV &          0.131 &          0.383 &          1.189 &          4.098 & \textbf{864.5} &          0.594 &          1.122 &          1.841 &          5.830 &          777.7 \\
      Wav2CLIP &          0.176 &          0.457 &          0.737 &          2.636 &          998.6 &          0.306 &          0.708 &          1.170 &          4.037 &          838.6 \\
       perfect &         98.404 &         99.833 &         99.966 &        100.000 &            1.0 &        100.000 &        100.000 &        100.000 &        100.000 &            1.0 \\
        random &          0.087 &          0.287 &          0.523 &          2.406 &         1005.9 &          0.110 &          0.332 &          0.617 &          2.878 &          891.0 \\

\midrule
&\multicolumn{10}{c}{CATEGORY} \\
\midrule
    LAION CLAP &          0.060 &          0.309 &          0.826 &          5.554 &          178.5 &          19.332 & \textbf{41.813} & \textbf{60.047} & \textbf{100.000} & \textbf{11.2} \\
  TimbreCLIP * &          0.076 &          1.023 &          1.475 &          5.128 &          243.6 & \textbf{22.073} &          39.893 &          54.894 & \textbf{100.000} &          11.9 \\
TimbreCLIP ALV & \textbf{1.280} & \textbf{1.856} & \textbf{2.298} & \textbf{8.027} & \textbf{158.3} &          15.385 &          31.328 &          50.192 & \textbf{100.000} &          12.7 \\
      Wav2CLIP &          0.014 &          0.125 &          0.398 &          1.633 &          261.7 &           6.772 &          18.978 &          36.612 & \textbf{100.000} &          15.7 \\
       perfect &         18.441 &         32.741 &         47.588 &         80.885 &            1.0 &         100.000 &         100.000 &         100.000 & \textbf{100.000} &           1.0 \\
        random &          0.102 &          0.303 &          0.566 &          2.301 &          164.6 &           6.549 &          19.544 &          35.739 & \textbf{100.000} &          16.0 \\
\bottomrule
\end{tabular}
\caption{Evaluation of TimbreCLIP on a cross-modal retrieval task performed on synthesizer patches from the surge dataset}
\end{table*}
We evaluate TimbreCLIP with cross-modal retrieval on a set of synth patches. In cross-modal retrieval, we want to find relevant documents from a set of documents using queries specified in a different modality. We achieve this by projecting our query and documents into a shared latent space and then computing a distance between the query and each document. This distance is then used to determine whether a document is relevant to the query.
We look at two cross-modal retrieval tasks. The first task is text to synth patch retrieval. In this task, the query is a text string and the documents are synth patches. We construct three types of simulated text queries: patch title, patch category, and the concatenation of patch title and patch category. For example, a patch with the title ``Blue smile" and category ``Keys" would produce three queries: ``Blue smile", ``Keys" and ``Blue smile Keys". Each synth patch has multiple audio files, each corresponding to a different MIDI pitch. Given text $t$ and audio files $x_0,x_1,..x_n$, we define the distance between the text $t$ and the patch as $min(dist(z_{x_0},z_{x_1}...z_{x_n})$.
The second task is audio to text retrieval. In this task, we want to find text documents that are relevant to a query consisting of audio from a single note.

The ``surge dataset'' \cite{turian_one_2021} consists of musical notes from 2084 synth patches, all recorded with MIDI velocity 64. We only use the notes with MIDI pitches 24, 36, 48, 60, 72, 84, 96, 108. Some patches have indices as part of their title to indicate that they belong to a series of patches with the same name. We discard these indices. After discarding indices,there are 1783 unique titles and 1816 unique title category combinations. The patches each belong to one of 31 categories.

We evaluate two different TimbreCLIP models. One is trained on ALV and NSynth (TimbreCLIP *) and one is trained exclusively on ALV (TimbreCLIP ALV). We compare these with two models trained on general audio 1) Wav2CLIP, a model trained on audio-image pairs from around 200k 10-second video clips \cite{wu_wav2clip_2022}. This model is the model from which our TimbreCLIP models are finetuned. ; 2) LAION CLAP, A recently released model trained on around 630k text-audio pairs, obtaining state of the art results on text-audio retrieval and zero shot audio classification. We also include the performance of a random baseline averaged over 100  runs (random) and the theoretical optimal value for each metric (perfect). We report retrieval at K (R@k) for $k\in\{1,5,10,50\}$ and average rank of the first relevant document (RANK) for both text to patch and audio text retrieval .

We notice that the recall@K values for the tasks involving the patch titles are considerably lower than what is typically seen on text-audio retrieval tasks.\cite{wu_large-scale_2022} This is perhaps an indication that matching patch titles with patch audio is a harder task than typical cross-modal retrieval on general audio. We reserve further analysis of these results to future work.
\section{Applications of TimbreCLIP}

\subsection{Text-guided Audio Equalization}
\begin{figure}[t]
\centering
\includegraphics[width=0.95\columnwidth]{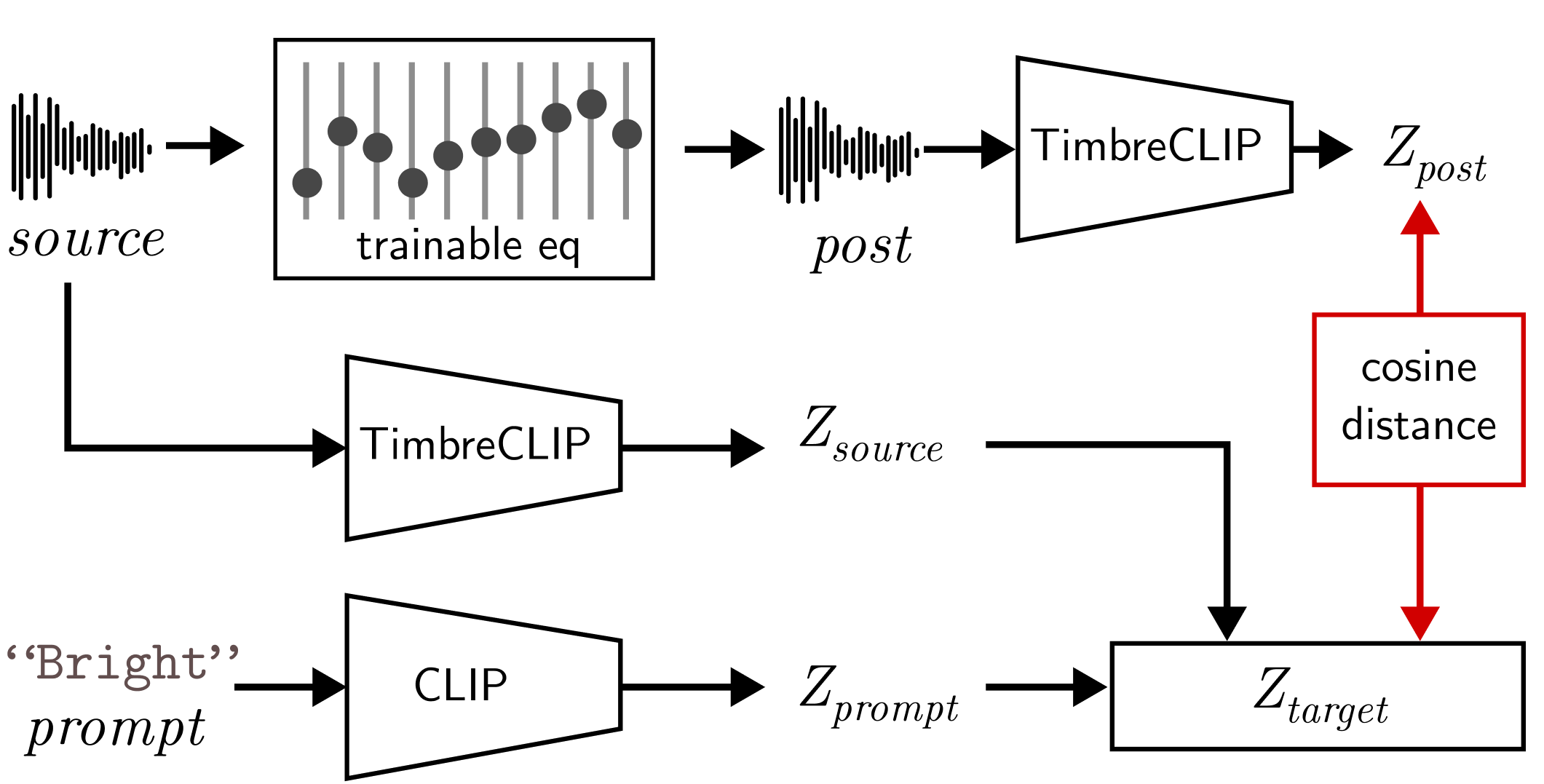} 
\caption{\textbf{Text-guided automatic audio equalization.} The source audio is encoded to form $z_{source}$. The prompt is encoded to form $z_{prompt}$. We then compute a target embedding $z_{target}=\alpha_{source} z_{source}+ \alpha_{prompt} z_{prompt}$. Each iteration, the source is processed by the trainable EQ and the processed output is encoded as $z_{post}$. The parameters of the trainable EQ are tuned with gradient descent so as to minimize the loss $dist(z_{post},z_{target})$. We optimize the EQ for 5k iterations using the Adam optimizer with a learning rate of 1e-2. We can also mix multiple text prompts to form our target $z_{target}=\alpha_{source}z_{source}+\alpha_{prompt_1}z_{{prompt}_1} + ... + \alpha_{prompt_n} z_{prompt_n}$.
}
\label{eq_diagram}
\end{figure}

\begin{figure*}
    \centering
    \includegraphics[width=\textwidth]{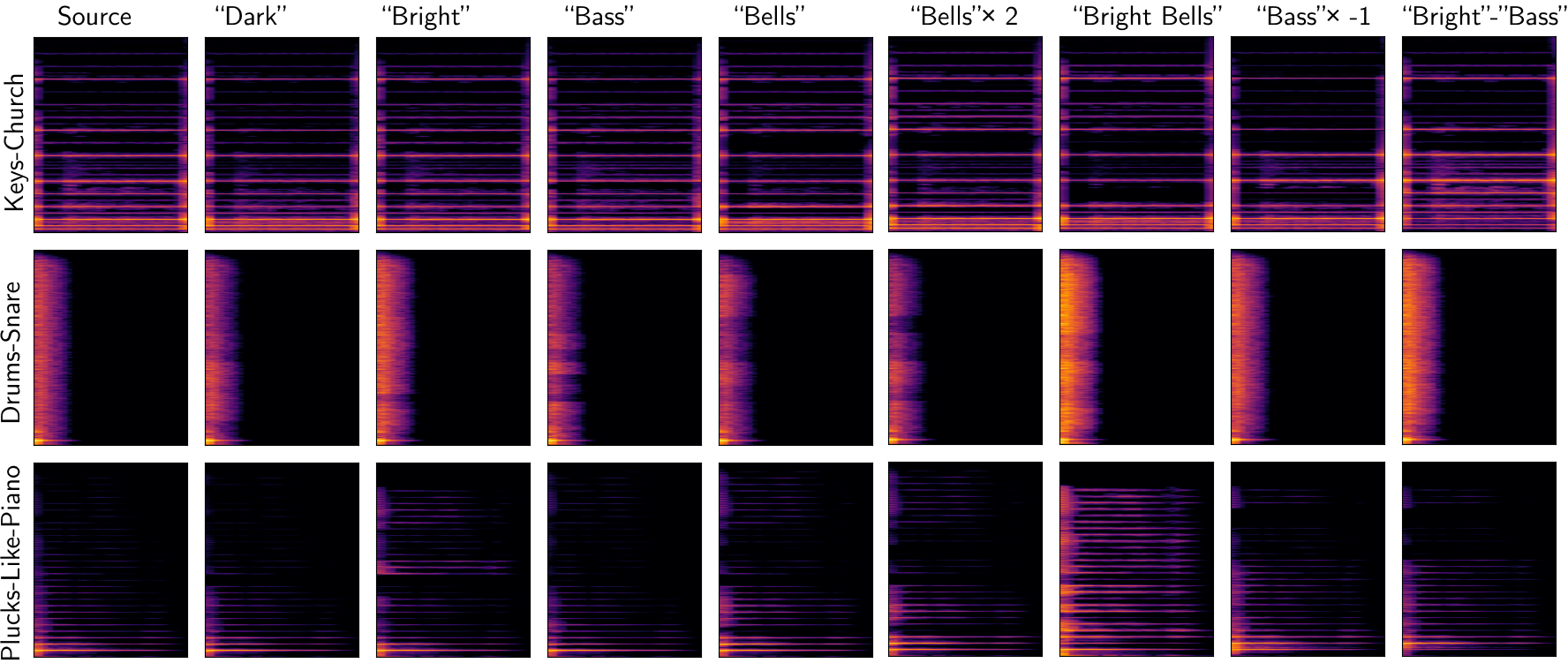}
    \caption{Spectrograms of 3 different sounds (rows) from the surge dataset before and after text-driven audio equalization with 8 different prompts (columns). Accompanying audio is available in the web supplement.}
    \label{fig:eq_spec}
\end{figure*}

An equalizer (EQ) is an signal processing module which amplifies or attenuates frequency bands of an audio signal according to controls set by the user. 
We demonstrate text driven audio equalization using TimbreCLIP. Figure \ref{eq_diagram} explains how the method works. Examples are shown in Figure \ref{fig:eq_spec}.

\subsection{Timbre to Image Generation}

\begin{figure}
    \centering
    \includegraphics[width=0.95\columnwidth]{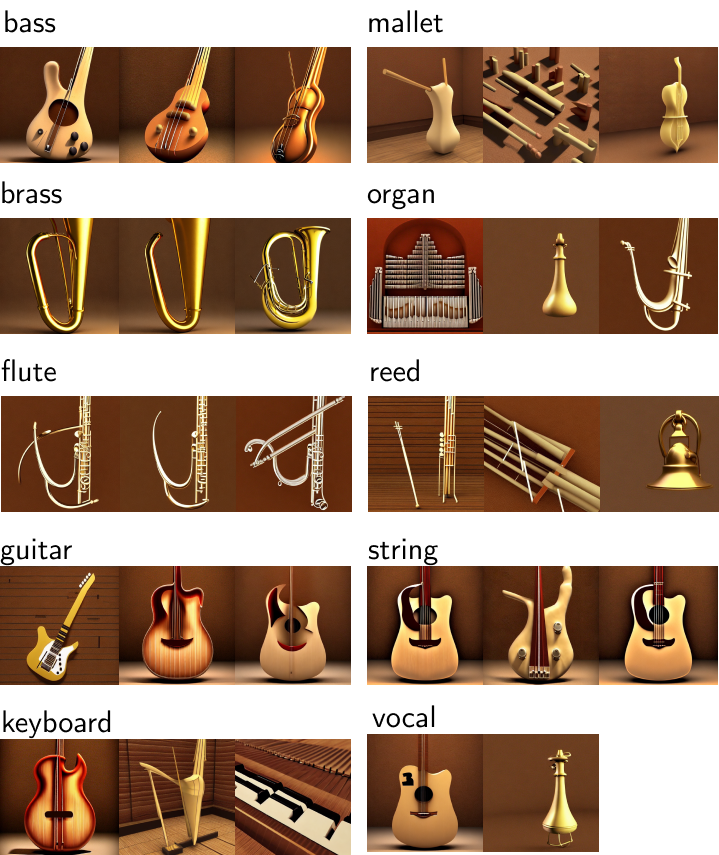}
    \caption{29 instrument sounds from 10 families from the NSynth test set visualized using TimbreCLIP and Stable Diffusion with prompt embedding interpolation. The template is: \texttt{A 3d render of a \textlangle keyword\textrangle, trending pinterest aesthetic}. The keywords used are names of 27 musical instruments. Audio and keywords used is available in the web supplement.} 
    \label{fig:t2i_nsynth}
\end{figure}

\begin{figure}
    \centering
    \includegraphics[width=0.95\columnwidth]{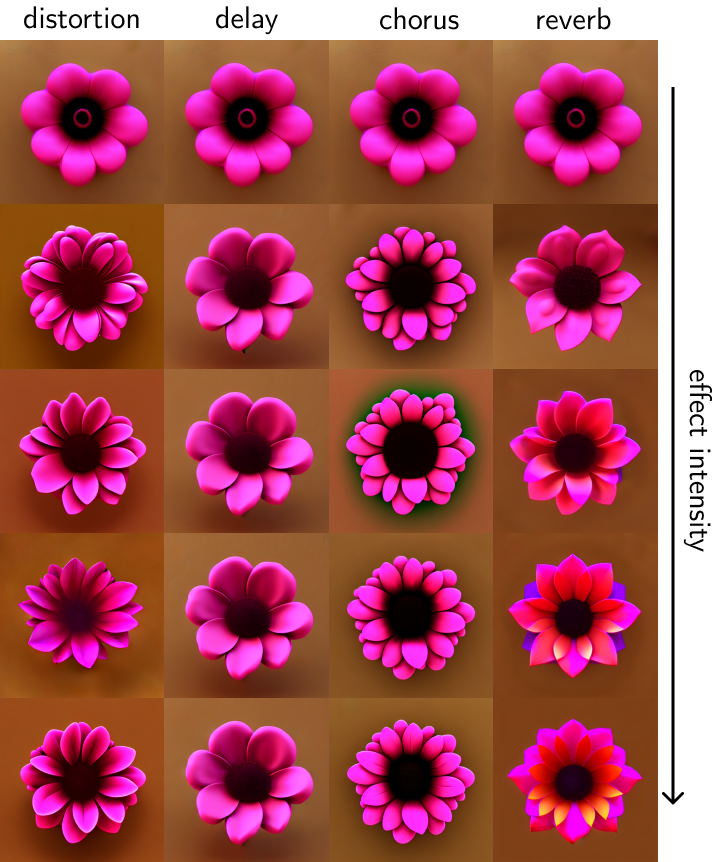}
    \caption{4 different audio effects visualized using TimbreCLIP and Stable Diffusion with prompt embedding interpolation. Each column corresponds to a sound effect and each row corresponds to a level of intensity of the effect that is applied. The template is: \texttt{A 3d render of a \textlangle keyword\textrangle flower, trending pinterest aesthetic}. The keywords are 12 adjectives.  Audio and keywords used is available in the web supplement.}
    \label{fig:t2i_effects}
\end{figure}

Stable diffusion \cite{rombach_high-resolution_2022} is an image generation system that can perform text to image synthesis. A text prompt $y$ is encoded with a text encoder $\tau$ whose output $t_y\in\mathbf{R}^{M \times d_\tau}$ is in turn used as conditioning for the image generation.
Anecdotal evidence suggests that interpolating between different $t$ generated from different prompts result in images that blends the concepts from the prompts.\footnote{\url{https://keras.io/examples/generative/random_walks_with_stable_diffusion/}} We propose to generate images from instrument audio by using a weighted average of embeddings of prepared prompts. We start by defining a set of keywords $k_0,k_1..k_n$. We then construct prompts $p_0,p_1..p_n$ from these keywords by injecting them into a prompt template such as \texttt{"A \textlangle keyword\textrangle  flower"}. We then embed each constructed prompt with $\tau$ to obtain prompt embeddings $t_0,t_1..t_n$. We also compute keyword embeddings $z_0,z_1..z_n$ using the CLIP text encoder.
To turn audio $x$ into an image, we first encode the audio file using our TimbreCLIP audio encoder $a(\cdot)$ to obtain our audio embedding $z = a(x)$. We then compute our image generation conditioning embedding with the following formula $t=dist(z,z_0)t_0+dist(z,z_1)t_1+..+dist(z,z_n)t_n$. We can also manipulate the weight distribution of each prompt by applying a softmax with a temperature parameter across the source keyword distances. Figure \ref{fig:t2i_nsynth} shows visualisations of instruments from the NSynth test split. We can also visualize an audio effect $f$ by instead setting $z$ as follows: $z=a(f(x))-a(x))$. Figure \ref{fig:t2i_effects} contains visualisations of various audio effects with varying intensity.

\section{Future work}
We plan to improve on the current TimbreCLIP models. We also plan to further develop our evaluation methodology. We will also look into potential ethical implications of this work. Finally, we aim to apply TimbreCLIP to more applications such as guiding synthesizers with text. 


\clearpage
\bibliography{references.bib}
\end{document}